\documentclass[jpa,referee]{svjour}
\usepackage{graphics}
\usepackage{epsfig}
\usepackage{amssymb}
\usepackage{amsfonts}
\usepackage{amsmath}
\usepackage{graphicx}
\newcommand\rd{{\rm d}}

\newcommand\NN{{\hbox{I\kern-.14em{N}}}}
\newcommand\RR{{\hbox{I\kern-.14em{R}}}}
\newcommand\ZZ{{\hbox{I\kern-.14em{Z}}}}

\newtheorem{theo}{Theorem}
\newtheorem{cor}{Corollary}

\def\boxit#1{\vbox{\hrule \hbox{\vrule
      \kern12truept\vbox{\kern12truept#1\kern12truept}\kern12truept
    \vrule}\hrule }}

\begin{document}

\title{Conservation of writhe helicity under anti-parallel reconnection\\}
\author{Christian E. Laing\inst{1}, Renzo L. Ricca\inst{2} \and De Witt L. Sumners\inst{3}
\thanks{\emph{Corresponding author:} renzo.ricca@unimib.it 
}%
}                     
\institute{Dept. Mathematics \& Computer Science, Dept. Biology, Wilkes University\\
84 West South St., Wilkes Barre, PA 18766, USA\\
E-mail: christian.laing@wilkes.edu
\and
Department of Mathematics \& Applications, U. Milano-Bicocca\\ 
Via Cozzi 55, 20125 Milano, Italy\\
E-mail: renzo.ricca@unimib.it
\and
Department of Mathematics, Florida State University\\
1017 Academic Way, Tallahassee, FL 32306-4510, USA\\
E-mail: sumners@math.fsu.edu}
\date{Submitted: \today\\}
\maketitle
\begin{abstract}
Reconnection is a fundamental event in many areas of science, from the  
interaction of vortices in classical and quantum fluids, and magnetic flux 
tubes in magnetohydrodynamics and plasma physics, to the recombination in 
polymer physics and DNA biology. By using fundamental 
results in topological fluid mechanics, the helicity of a flux tube can 
be calculated in terms of writhe and twist contributions. Here we show that 
the writhe is conserved under anti-parallel reconnection. Hence, for a pair 
of interacting flux tubes of equal flux, if the twist of the reconnected tube 
is the sum of the original twists of the interacting tubes, then helicity is 
conserved during reconnection. Thus, any deviation from helicity conservation 
is entirely due to the intrinsic twist inserted or deleted locally at the 
reconnection site. This result has important implications for helicity and 
energy considerations in various physical contexts.
\end{abstract}

\vspace{1cm}
\noindent \textbf{Keywords:} reconnection; recombination, flux tubes; vortex tubes; 
helicity; linking number; writhe; twist; topological dynamics; structural complexity

\newpage

\begin{figure}[t]
\begin{center}  
\includegraphics[width=0.8\textwidth]{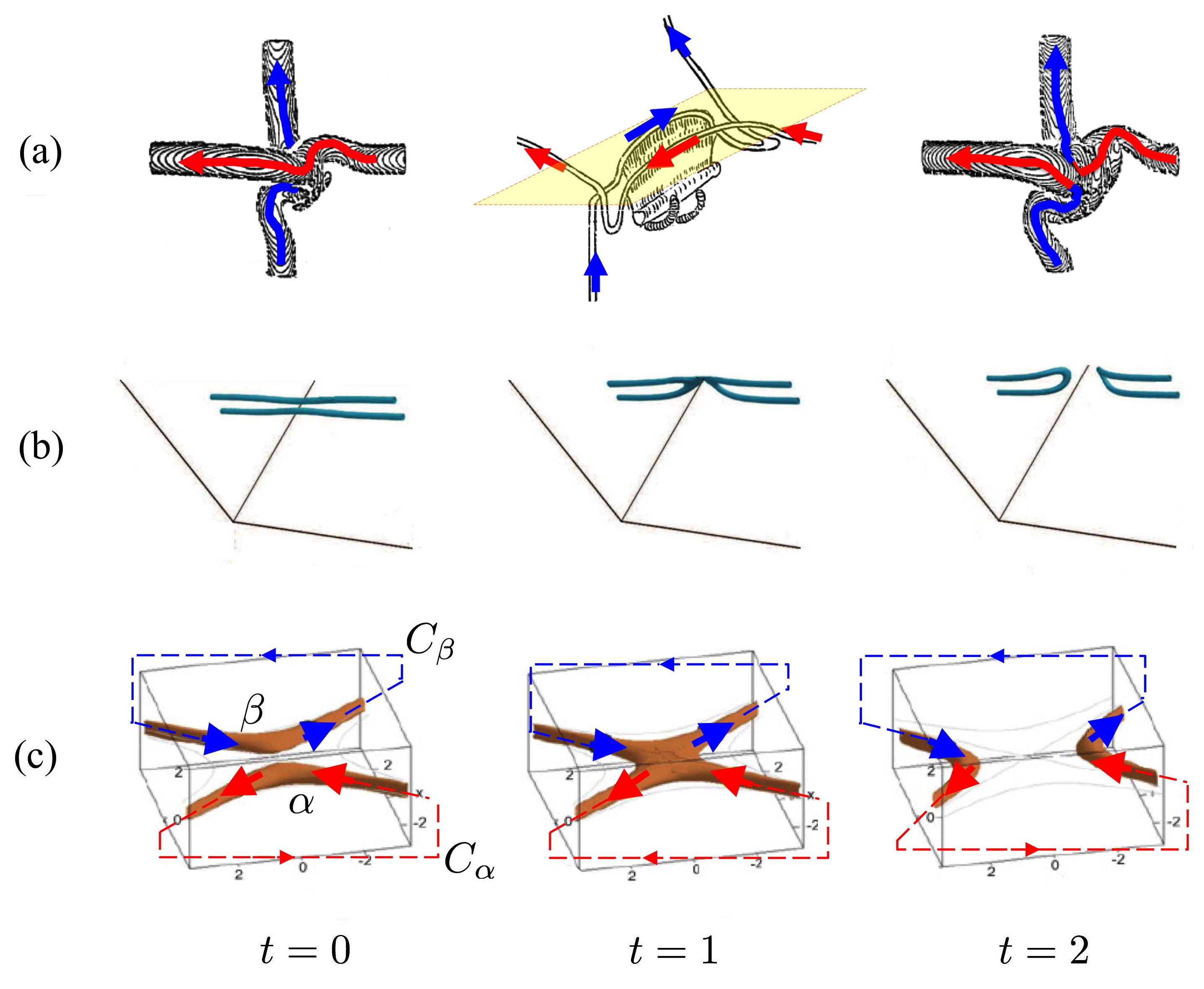}
\caption{}
\label{FIGURE1}
\end{center}
\end{figure}

\noindent
{\Huge F}ilamentary structures, such as vortex filaments in classical and quantum 
fluids,${}^\textrm{\cite{KT94,BPSL08,HD11,ZCBB12}}$ 
magnetic flux tubes,${}^\textrm{\cite{LF96,PF00}}$ 
phase defects,${}^\textrm{\cite{L99}}$ and polymers and 
macromolecules${}^\textrm{\cite{Sumners95,VCLPKDS98}}$ 
are ubiquitous in nature. When parts of these 
filaments come sufficiently close to one another, they tend to influence each other and 
recombine through reconnections (see Fig.~\ref{FIGURE1}). Reconnection is a process associated 
with a change of topology and geometry of the interacting filaments by an exchange of the neighboring 
strands.${}^\textrm{\cite{GMP12}}$ In general, when two disjoint, closed tubes (like vortex rings) reconnect, 
the result is a single closed tube and when a single closed tube reconnects with itself, the result is 
two closed tubes.  Such a morphological change is typically accompanied by a change in energy, 
partly dissipated due to small-scale effects associated with viscosity, resistivity or other.  
Thus, detailed study of reconnections is crucial to understand energy re-distribution and dissipation 
in many fluid systems, from vortex tangles in classical and superfluid 
turbulence,${}^\textrm{\cite{vRea12,K13,ZCBB12}}$ to phase transitions in mesoscopic 
physics,${}^\textrm{\cite{L99}}$
from astrophysical flows in solar and stellar physics${}^\textrm{\cite{PF00,CDS11}}$ to 
confined plasmas in fusion physics.${}^\textrm{\cite{MS88,B00}}$
Detailed analysis based on direct numerical simulations of real fluid equations 
reveals certain qualitative common features of the reconnection event (compare for instance the 
various scenarios shown in Fig.~\ref{FIGURE1}). In the majority of cases at the time of closest approach  
the interacting tubes tend to align themselves in an anti-parallel 
fashion, followed by a reconnection of the local strands through a rapid, merging process in a direction 
orthogonal to their mutual alignment before final separation. Fine details of the 
reconnection event (such as the generation of secondary, bridge structures in vortex dynamics) may differ 
from case to case, but certain geometric features such as anti-parallel alignment of the reconnecting strands 
and transversal merging seem to have a generic character. Qualitatively similar features, for istance, seem 
to characterize recombination events in polymer physics as well as 
in DNA biology,${}^\textrm{\cite{Sumners95,Vazquez04}}$ when two unknotted circular DNA plasmids 
are joined into a single plasmid in a site-specific recombination 
event.${}^\textrm{\cite{Stasiak96,VCLPKDS98,CWPSC99,Weber13}}$
These common geometric features are the focus of this paper. 

\noindent\textbf{\large Results}\\
\noindent
\textbf{Helicity, linking numbers and writhe.}
In fluid systems a fundamental quantity, that detects topological information and that has a strict 
relation with energy, is the \textit{helicity} $H$ of fluid flows (kinetic or magnetic). 
For two interacting disjoint tubular filaments $\alpha$ and $\beta$, centered on their respective curves 
$C_\alpha$ and $C_\beta$ (see Fig.~\ref{FIGURE1}c), the helicity $H=H(\alpha,\beta)$ can be 
written as${}^\textrm{\cite{M69,BF84,MR92,RM92}}$
\begin{equation}
  H(\alpha,\beta)=\Phi_\alpha^2 SL(\alpha)+\Phi_\beta^2 SL(\beta) 
  +2\Phi_\alpha\Phi_\beta Lk(C_\alpha, C_\beta)\ ,
\label{HL} 
\end{equation}
where $\Phi$ is a measure of the tube flux (field strength), and $SL$ and $Lk$ are topological 
numbers denoting self-linking and mutual linking of the two flux-tubes, respectively (for 
their definitions see Refs.~\cite{Calu61}, \cite{Pohl68}, \cite{White69}, and text below). 
During reconnection, the interacting tubes may change strength, whereas topology certainly changes; 
hence a change in helicity should be expected. Even when the flux remains conserved (as in the case 
of quantized vortices in superfluid helium), a change in linking numbers may happen, because the 
reconnection of a pair of closed, oriented curves produces a single closed, oriented curve 
(with no linking number), and vice versa. Here all curves are tacitly assumed  
to be smooth, with the exception of the polygonal curves referred in the text below and in the next 
subsection. Polygonal curves are used to facilitate the proof of conservation of writhe under 
reconnection (since polygonal curves can approximate smooth curves arbitrarily closely).
Since reconnection is a local process, the morphological and structural change experienced by 
the reconnecting strands is reflected in the change of the individual self-linking numbers. 
For a single flux tube $\alpha$, $SL(\alpha)$ admits decomposition  
into two geometric quantities, the writhe $Wr(C_\alpha)$ of the tube centerline $C_\alpha$ and 
the twist $Tw(R_\alpha)$ of the tube reference ribbon 
$R_\alpha$;${}^\textrm{\cite{Fuller71}}$
from standard differential geometry, the twist can be decomposed into two parts, given 
by the normalized total torsion $T(C_\alpha)$ of $C_\alpha$, and the intrinsic twist 
$N(R_\alpha)$ of $R_\alpha$ around $C_\alpha$.  Thus, we have 
\begin{equation}
  SL(\alpha)= Wr(C_\alpha)+Tw(R_\alpha)=Wr(C_\alpha)+ T(C_\alpha)+N(R_\alpha)\ . 
\label{SLalpha} 
\end{equation}
Since writhe and twist are geometric quantities, their values change continuously with the continuous 
change in space of the curve $C_\alpha$ and the reference ribbon $R_\alpha$. 

Writhe is a geometric measure of \textit{non-planarity} for spatial 
curves;${}^\textrm{\cite{Fuller71,LS08}}$ indeed, planar curves and closed curves on a round 2-sphere have 
\textit{zero} writhe. 
Let the unit sphere $S^2$ denote the space of directions (unit vectors) in $\mathbb{R}^3$.  
Given an oriented, simple, closed curve $A$ in $\mathbb{R}^3$, consider a generic planar 
projection (knot diagram) of $A$ in the direction $\nu\in S^2$, with standard sign convention 
of $\pm 1$ for over/under--passes. One now adds up all of the signed crossings to obtain the 
\textit{directional writhe} of $A$, $\omega_\nu(A)$.  By averaging the directional writhe 
over all directions, one obtains the \textit{writhe} of $A$:
\begin{equation}
Wr(A)=\frac1{4\pi}\sum_{\nu\in S^2}{\omega_\nu(A)}\,.
\label{writheA}
\end{equation}
Given a pair of disjoint, simple, closed curves $\{A,B\}$, the \textit{linking number} 
$Lk(A,B)$ can be calculated from any generic projection of the pair of curves by adding up 
the crossings between the curves (neglect the self-crossings of each curve) as 
follows.  Suppose that there are $n$ crossings $\{X_i,1\le i \le n\}$ between $A$ and $B$, 
and $\epsilon_i=\pm1$ denotes the sign of the $i$-th crossing according as the crossing is 
positive or negative, then we have   
\begin{equation}
Lk(A,B)=\frac12\sum_{i=1}^n \epsilon_i\,.
\label{LkAB}
\end{equation}
Since the linking number is constant over all projections, averaging the value 
over all projections does not change this value.

Suppose now that $A$ is an oriented $n$-edge polygon with edges $\{a_i,1 \le i \le n\}$, and 
$B$ is an $m$-edge polygon with edges $\{b_j, 1 \le j \le m\}$.  Consider a pair of distinct oriented 
edges $\{a_i, a_j\}$ of $A$.  We wish to compute the contribution to the writhe of $A$ from the pair 
of edges $\{a_i, a_j\}$.  The set of all directions on $S^2$, where one sees a single crossing between 
these edges, is an open set; moreover, one sees the same crossing sign over this entire open set. 
Under the antipodal map on $S^2$, a map that takes any point $x\in S^2$ to $-x$, this open set is 
invariant, since a crossing seen in a given direction is seen as a crossing of the same sign in the 
opposite direction.  The contribution to the writhe of $A$ from the pair of edges $\{a_i, a_j\}$ 
is $\omega(a_i, a_j)$, the signed area on the unit 2-sphere $S^2$ of this open set.  Note that 
$\omega(a_i, a_j)=0$ if $i=j$, or if the edges meet in a common vertex --- in each case the edges 
are identical or co-planar, with no crossings visible under any projection direction. 
We can compute $Wr(A)$ in terms of the edges of polygon $A$:
\begin{equation}
Wr(A)=\frac1{4\pi}\sum_{i=1}^{n}\sum_{j=1}^{n}\omega(a_i,a_j)\,.
\label{writheAedges}
\end{equation}
For disjoint oriented polygons $\{A,B\}$, we can compute $Lk(A,B)$ in terms of the edges
\begin{equation}
Lk(A,B)=\frac1{4\pi}\sum_{i=1}^{n}\sum_{j=1}^{m}\omega(a_i,b_j)\,,
\label{LkABedges}
\end{equation}
and similarly the writhe of the disjoint union of $A$ and $B$:
\begin{eqnarray}
Wr(A\cup B) &=& \frac1{4\pi} \bigg[\sum_{i=1}^{n}\sum_{j=1}^{n}\omega(a_i,a_j)+
  \sum_{i=1}^{n}\sum_{j=1}^{m}\omega(a_i,b_j)\nonumber\\[3mm]
  &&+\sum_{j=1}^{m}\sum_{i=1}^{n}\omega(b_j,a_i) +\sum_{i=1}^{m}\sum_{j=1}^{m}\omega(b_i,b_j)
  \bigg]  \nonumber\\[3mm]
  &=& Wr(A)+2Lk(A,B)+Wr(B)\,.
\label{writheAUB} 
\end{eqnarray}

\begin{figure}
\begin{center}  
\includegraphics[width=0.7\textwidth]{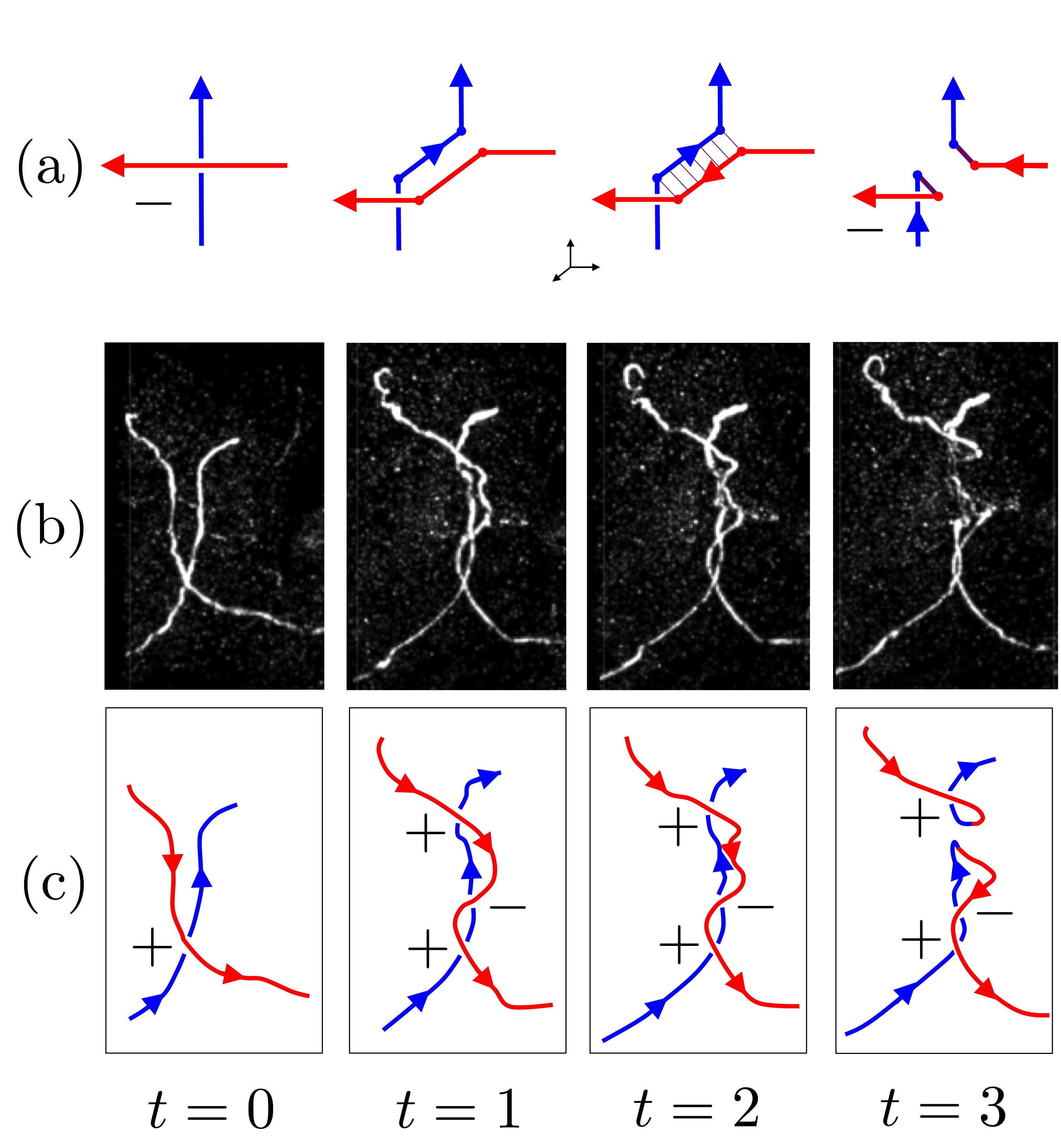}
\caption{}
\label{FIGURE2}
\end{center}
\end{figure}

\noindent
\textbf{Reconnection conserves writhe.}
Experimental and computational evidence shows that reconnection is a process that 
takes place along the interacting segments of two tube centerlines 
(see Fig.~\ref{FIGURE2}b), and does not occur at a point in isolation. Hence, when 
the interacting segments of two tubes approach each other, the reconnection event 
can only take place \textit{near} an apparent crossing point (and not \textit{at} a crossing 
point, that in any case depends on the projection direction). Directional writhe, on the 
other hand, depends on the projection, and only when it is averaged over all directions 
of sight it becomes a projection independent measure (as in eq. (\ref{writheA})). Thus, 
reconnection near a crossing does not change the writhe (see Fig.~\ref{FIGURE1}a). 
Fig.~\ref{FIGURE2}b shows close up screen shots of the anti-parallel 
alignment of two trefoil vortex strands and subsequent reconnection. From direct
inspection of the supplementary material made available by \textit{Nature Physics},
we can see (from the smooth tracings of Fig.~\ref{FIGURE2}c) that   
the red vortex line has been moved across the top of the blue vortex line ($t=0,1$) and 
then the anti-parallel reconnection segments are spatially juxtaposed ($t=2$). 
The configuration just after reconnection is shown in $t=3$.  The directional writhe in 
each of the figures at $t=0,1,2,3$ is $+1$.  This reconnection event is very fast compared 
with the typical vortex evolution time, so that the writhe of the unseen rest of the 
configuration remains essentially constant throughout this quick reconnection.  
Although we only have one projection direction shown in the screen shots, the pair of 
vortex segments are very close to co-planar just before and just after reconnection takes 
place, so the directional writhe is very close to the true writhe. In this experiment, 
we see that observed reconnection of the trefoil vortex to the Hopf link vortex conserves writhe.

A rigorous proof that anti-parallel reconnection conserves writhe is given here below. 
Our result will not depend on any specific projection and proof relies on the following 
assumptions: 

\noindent 
A1: under reconnection, orientation is preserved; \\
A2: the reconnecting segments are oriented in an anti-parallel fashion; \\ 
A3: the reconnecting segments are \textit{isomorphic}, identical under spatial translation.\\

\begin{figure}[t]
\begin{center}  
\includegraphics[width=1\textwidth]{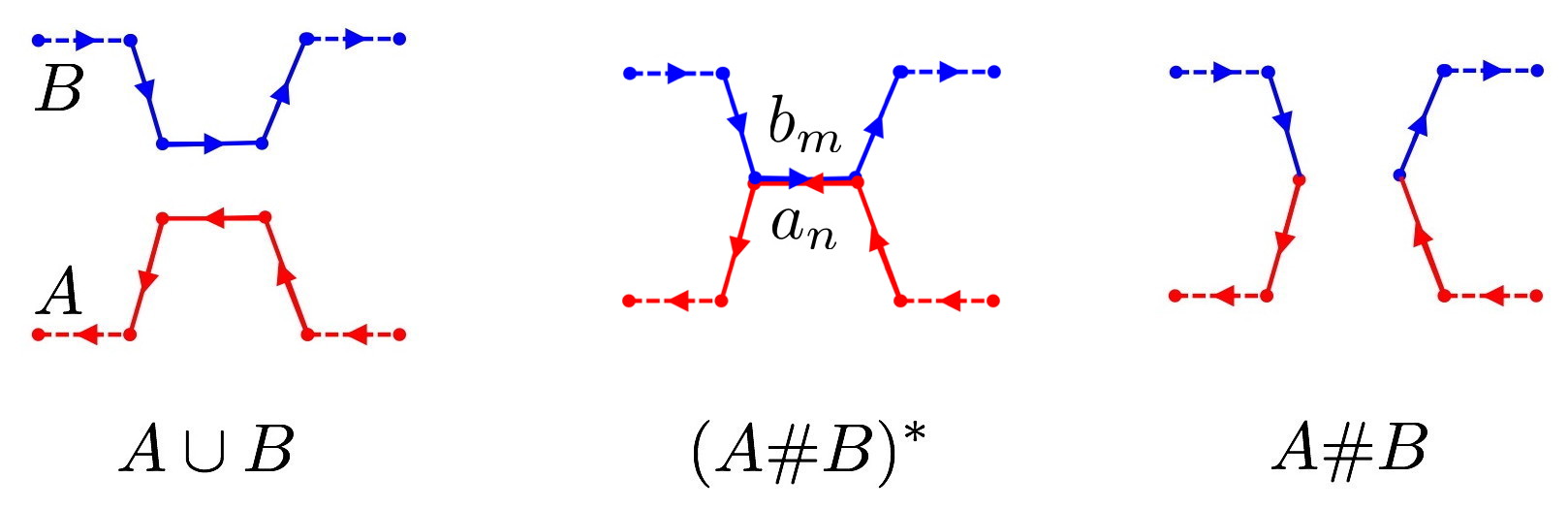}
\caption{}
\label{FIGURE3}
\end{center}
\end{figure}

Now, suppose that we have two disjoint oriented polygons $A = \{a_i, 1 \le i \le n\}$ and 
$B = \{b_j, 1 \le j \le m\}$, that have the following properties:

\noindent
(i) edges $a_n$ and $b_m$ have the same length;\\
(ii) polygon $B$ can be translated without intersecting polygon $A$ until the edges 
$a_n$ and $b_m$ are coincident with opposite orientation (as in the central diagram of 
Fig.~\ref{FIGURE3}).

When edges $a_n$ and $b_m$ are coincident, one has formed the $\theta$-curve intermediate 
$(A\#B)^\ast$; by deleting the interior of the common edge $a_n = b_m$ from $(A\#B)^\ast$, 
one obtains the oriented reconnected curve $(A\#B)$.   

Consider the effect of the translation that aligns $b_m$ with $a_n$ on each of the terms in 
equation (\ref{writheAUB}) for $Wr(A\cup B)$:  since translation is a rigid motion, $Wr(A)$ 
and $Wr(B)$ are unchanged during the translation, and $2Lk(A,B)$ is a topological invariant 
unchanged by translation. At the end of translation, when $a_n = b_m$, if we stipulate that 
in the calculation of $Wr[(A\#B)^\ast]$ we will count the common edge $a_n = b_m$ twice 
(with opposite orientations for $a_n$ and $b_m$), then we have shown
\begin{equation}
Wr(A\cup B)=Wr[(A\#B)^\ast]\,.
\label{WrAUB}
\end{equation}
Since $a_n = b_m$ with opposite orientations, for each edge $e$  in  $A\cup B$, we have  
$\omega(a_n,e) = -\omega(b_m,e)$, so in the calculation for $Wr[(A\#B)^*]$ these terms cancel out in 
pairs, and we are left with the writhe of the reconnected curve $(A\#B)$, and we have proved:
\vskip4pt plus2pt
\noindent
\begin{theo}
\label{theorem1}
Reconnection conserves writhe: for disjoint oriented polygons $A$ and $B$ (satisfying properties
(i) and (ii) above), $Wr(A\cup B)=Wr[(A\#B)]$.
\end{theo}
\vskip4pt plus2pt
When a single curve reconnects with itself to produce a pair of curves, the writhe of the single curve 
may change as the reconnection segments are aligned and brought into spatial juxtaposition.  However, 
as the segments to be juxtaposed are moved closer and closer together, the writhe of the configuration 
approaches a limiting value, the writhe of the theta-curve intermediate. This limiting value of the writhe is 
equal to the writhe of the reconnected pair of disjoint curves.

\begin{figure}
\begin{center}  
\includegraphics[width=0.8\textwidth]{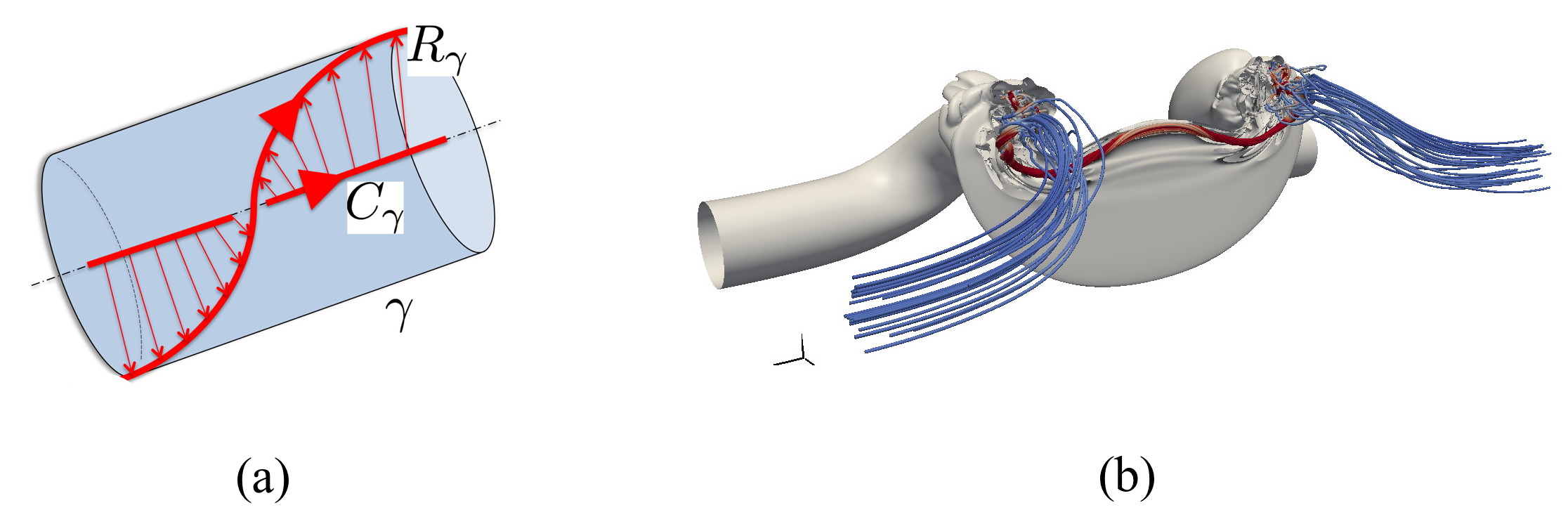}
\caption{}
\label{FIGURE4}
\end{center}
\end{figure}

\vspace{0.5cm}
\noindent
\textbf{Conservation of helicity under anti-parallel reconnection.}
Fig.~\ref{FIGURE4}a  shows the flux tube $\gamma$, with center curve $C_\gamma$ and 
flux ribbon $R_\gamma$, formed by connecting $C_\gamma$ with one of the field lines in 
$\gamma$.  Suppose also that flux tube $\gamma$ has flux $\Phi$. For a single flux tube 
$\gamma$ eqs. (\ref{HL}) and (\ref{SLalpha}) give us
\begin{equation}
H(\gamma)=\Phi^2[Wr(C_\gamma)+Tw(R_\gamma)]\ .
\label{helicity}
\end{equation}
By using the right-hand side decomposition given by eq. (\ref{SLalpha}), we can 
distinguish the \textit{centerline helicity} $H_C=\Phi^2[Wr(C_\gamma)+T(C_\gamma)]$, 
that depends solely on tube axis geometry (so that can be entirely estimated by 
external measurements of $C_\gamma$), from the \textit{intrinsic twist helicity} 
$H_N=\Phi^2 N(R_\gamma)$, that depends on the internal twist of the field line 
distribution. Let $T(s)$ denote the unit tangent vector 
at position $s$ on the curve $C_\gamma$ (parameterized by arc length $s$), 
and $V(s)$ denote a unit normal vector pointing from $C_\gamma$ to 
the edge of ribbon $R_\gamma$ at position $s$.  The \textit{incremental twist} of 
the ribbon $R_\gamma$ along 
the center line $C_\gamma$ (in the direction of $T$) at position $s$ is given by 
$w(s)=(\frac{dV}{ds}\times V)\cdot T$ (see Refs.~\cite{Fuller71}, \cite{MR92}, \S 5).  
The \textit{total twist} is thus given by the line integral: 
\begin{equation}
Tw(R_\gamma)=\frac1{2\pi}\int_{C_\gamma}w(s)\,\rd s
=\frac1{2\pi}\int_{C_\gamma}\left(\frac{dV}{ds}\times V\right)\cdot T\,\rd s\ . 
\nonumber
\label{Tw}
\end{equation}
Suppose now that we have two disjoint flux tubes $\{\alpha, \beta\}$ with equal flux $\Phi$.  
Take $\Phi  = 1$  for simplicity.  Suppose also that the oriented center lines 
of tubes $\alpha$  and $\beta$ satisfy the smooth version of conditions (i) and (ii) of 
Theorem~\ref{theorem1} above for reconnection.  Specifically, center lines $C_\alpha$ and 
$C_\beta$  are each divided into two arcs: $C_\alpha = C_{\alpha 0} \cup C_{\alpha 1}$, and 
$C_\beta = C_{\beta 0} \cup C_{\beta 1}$. In the reconnection event, $C_\beta$ is translated (without 
crossing $C_\alpha$) until arcs $C_{\alpha 0}$ and $C_{\beta 0}$ are coincident (with opposite 
orientation), producing the $\theta$-curve intermediate ($C_\alpha \# C_\beta)^\ast$.  At this time, 
the (infinitesimally small) coincident arc $C_{\alpha 0} = C_{\beta 0}$  is removed, producing the 
reconnected curve  $C_\alpha \# C_\beta =  C_{\alpha 1} \cup C_{\beta 1}$. 
Before reconnection (see, for example, Fig.~\ref{FIGURE1}c), we have:
\begin{eqnarray}
  H(\alpha\cup\beta) &=& Wr(C_\alpha\cup C_\beta)+Tw(R_\alpha\cup R_\beta)\nonumber\\
  &=& Wr(C_\alpha) + Wr(C_\beta)+2Lk(C_\alpha,C_\beta)\nonumber\\
  &&+ Tw(R_\alpha)+Tw(R_\beta)\ . 
\label{helicitycup} 
\end{eqnarray}
Preliminary results along the lines of the last eq. (\ref{helicitycup}), based on linking numbers and 
mutual winding of magnetic lines (but not on writhe and twist decomposition), can be found in 
Ref.~\cite{MS88}. Since the ribbons $R_\alpha$ and $R_\beta$ are 
disjoint, then the twist of the union of the ribbons is the sum of the individual twists of each 
ribbon. Given that the flux tubes are locally aligned for reconnection, then translating $C_\beta$ to 
$C_\alpha$ conserves the individual twist integrals.  

For the $\theta$-curve intermediate, we assume that the superimposed arc $C_{\alpha_0} = C_{\beta_0}$ 
 has both ribbons on it, so the twist of this ribbon over the $\theta$-curve  intermediate 
$(C_\alpha \# C_\beta)^\ast$ has total twist the sum of the individual twists.  
The twist of the ribbon over the reconnected center 
lines $R(C_\alpha \# C_\beta)$ is $Tw(R_{\alpha_1}) + Tw(R_{\beta_1})$. We have the following 
equation for the change in twist due to reconnection: 
\begin{eqnarray}
  \Delta Tw&=&[Tw(R_\alpha)+Tw(R_\beta)]-[Tw(R_{\alpha_1}) +Tw(R_{\beta_1})]\nonumber\\
  &=&Tw(R_{\alpha_0})+Tw(R_{\beta_0})\ . 
\label{deltatwist} 
\end{eqnarray}
In a reconnection event suppose now that twist is conserved, i.e.
\begin{equation}
\Delta Tw=0\ .
\label{deltatwist0}
\end{equation}

Given this, we have conservation of helicity:
\vskip4pt plus2pt
\noindent
\begin{theo}
\label{theorem2}
Given anti-parallel reconnection of flux tubes $\{\alpha,\beta\}$ with equal flux $\Phi$, 
if the total twist of the flux tube ribbons is conserved, then helicity is also conserved, that is 
\begin{equation}
H(\alpha\cup\beta)=H(\alpha\#\beta)\,.
\label{Hab}
\end{equation}
\end{theo}
\vskip4pt plus2pt

\noindent
\textbf{Role of twist.}
Since the super-imposed edges have opposite orientation, 
it is possible that the line integrals over the edges have the same absolute value and different sign, giving 
us $\Delta Tw = 0$.  Moreover, the edges that get superimposed to form the $\theta$-curve intermediate can have 
vanishingly small length (or take the limit as the length of the super-imposed edge goes to zero).  At zero 
length (the $\theta$-curve intermediate now becomes a figure-of-eight, where $C_\alpha$ and $C_\beta$ have 
a vertex in common), the line integrals over the common vertex vanish, and $\Delta Tw = 0$.  This may be the case 
for reconnections of quantized vortex filaments in superfluids, whose typical vortex core cross-section is of the 
order of $10^{-10}$~m in Helium--4, several orders of magnitudes smaller than the 
average distance between vortices in typical laboratory experiments.${}^\textrm{\cite{BPSL08}}$. 
Furthermore, since a quantized vortex filament is essentially an empty cavity, we have no intrinsic twist, 
hence total twist reduces to total torsion (cf. eq. \ref{SLalpha}). Lack of internal structure, and hence of 
intrinsic twist, characterizes many other physical systems, such as atomic Bose--Einstein 
condensates,${}^\textrm{\cite{POB12}}$, phase line singularities in nonlinear optics${}^\textrm{\cite{ODP09}}$ 
and, possibly, superconductors,${}^\textrm{\cite{B09}}$ where reconnections may indeed 
trigger topologically complex structures. For all these systems any change in self-linking number 
(and helicity) should be ascribed to the sole change in total torsion through reconnection. 

As mentioned in the introduction (see again eq. \ref{SLalpha}), suppose that the smooth curve $C_\alpha$ is 
parameterized by arc-length $s$, and that $\tau(s)$ denotes the torsion at a point on the curve.  
The normalized total torsion $T(C_\alpha)$ of $C_\alpha$ is given by the integral
\begin{equation}
T(C_\alpha)=\frac1{2\pi}\int_{C_\alpha}\tau(s)\,\rd s\ .
\label{torsion}
\end{equation}
Suppose now that smooth curves $C_\alpha$ and $C_\beta$ are to be reconnected (in an anti-parallel fashion). 
The normalized total torsion of the reconnected curve is given by the integral 
\begin{eqnarray}
  T(C_\alpha \# C_\beta)&=&\frac1{2\pi}\left(\int_{C_\alpha}\tau(s)\,\rd s\ 
  + \int_{C_\beta}\tau(s)\,\rd s\ \right)\nonumber\\
  &=& \frac1{2\pi}\int_{C_\alpha \# C_\beta}\tau(s)\,\rd s\ . 
\label{tottorsion} 
\end{eqnarray}
Since for infinitesimally small, anti-parallel, co-planar arcs $T(C_{\alpha 0})=-T(C_{\beta 0})=0$
(total torsion is additive), we must have $T(C_\alpha \cup C_\beta)=T(C_\alpha \# C_\beta)$. 
Hence, 

\vskip4pt plus2pt
\noindent
\begin{cor}
\label{corollary}
If the intrinsic twist $N(R_{\alpha 0})\neq N(R_{\beta 0})$, then
\begin{equation}
\Delta H= H(\alpha \# \beta)-H(\alpha \cup \beta)=\Delta Tw=\Delta N\,.
\label{dh}
\end{equation}
\end{cor}
\vskip4pt plus2pt

Since total torsion is due to the contribution of the torsion of the tube axes over their entire length, a quantity 
that can be estimated or computed directly, any change in conformational energy through reconnection can be 
estimated via total torsion information quite accurately. When intrinsic twist is an important part of total 
twist (see Fig.~\ref{FIGURE4}b), careful considerations on the relative role of spatial gradients associated 
with curvature and torsion of the tube axis and intrinsic twist must be made. Since dissipative forces tend 
to erode higher order gradients first, it is natural to expect that, in general, $\Delta N \neq 0$. 
Hence, as a consequence of Theorem 2 above, any change in helicity should be ascribed to the sole 
change in intrinsic twist. 

\vspace{1cm}
\noindent\textbf{\large Discussion}\\
\noindent
We have proven that total writhe remains conserved under anti-parallel reconnection of flux 
tube strands. Since the helicity of a flux tube admits decomposition in terms of
writhe and twist, this result implies that for a pair of interacting flux tubes 
of equal flux, writhe helicity remains conserved throughout the reconnection process. In this case 
any deviation from helicity conservation is entirely due to the intrinsic twist inserted or 
deleted locally at the reconnection site. If the twist of the reconnected tube is the sum of 
the original twists of the individual tubes before reconnection, then the flux tube helicity 
is conserved during reconnection. 

The analogue of flux tube reconnection in molecular biology is site-specific recombination with 
directly repeated reconnection sites. The sites are oriented in anti-parallel alignment, and reconnection 
of a single DNA plasmid produces a pair of plasmids, and reconnection of a pair of plasmids produces a 
single plasmid. Recent very interesting work on the minimal DNA recombination pathway${}^\textrm{\cite{Setal13}}$ 
proves that if one starts with the trefoil, and insists that recombination reduces configuration 
complexity (minimal crossing number), then the minimal pathway trefoil $\to$ Hopf link $\to$  
unknotted circle $\to$ pair of unknotted, unlinked circles is exactly the 
reconnection pathway taken by the trefoil vortex in the Kleckner--Irvine 
experiment.${}^\textrm{\cite{KI13}}$.

Our result has therefore important implications well beyond fluid mechanics. For physical systems
where helicity and energy considerations are important, and in particular for magnetic fields in 
solar and plasma physics and for vortex flows in quantum 
and classical turbulence, reconnections are not only key to understand geometric and 
topological changes in the fluid flow structure,${}^\textrm{\cite{WB89,LF96,LDA01,DCB10,KI13}}$ 
but they are also responsible for crucial re-distribution and dissipation of the energy at smaller 
scales.${}^\textrm{\cite{SL89,FB93,vRea12,K13}}$. Our present results will help to address the 
focus of current research on the role of twist and on the finer details of the tube internal 
structure undergoing reconnection.


\vspace{0.5cm}
\noindent
\textbf{Acknowledgments}\\
In the concluding stages of this work we became aware of a parallel effort by Scheeler \textit{et al.} \cite{Sea14}, 
who independently identified the mechanism for conservation of "link plus 
writhe" through reconnections from their experimental observations, and tested its validity in experiments 
and simulations of topology changing vortex loops. 
The authors would like to express their gratitude to 
Keith Moffatt for his comments on a preliminary version of this manuscript.  The authors would also like 
to thank several mathematics institutes for support, during the preparation of this paper:  the Institute 
for Mathematics and Its Applications 
in Minneapolis, USA (CEL and DWS), the Centro di Ricerca Matematica Ennio De Giorgi in Pisa, Italy 
(RLR and DWS), and the Isaac Newton Institute for Mathematical Sciences in Cambridge, UK (RLR and DWS).

\vspace{0.5cm}
\noindent
\textbf{Author Contributions}\\
C.E.L., R.L.R. and DeW.L.S. have contributed equally in 
the preparation of this work.

\vspace{0.5cm}
\noindent
\textbf{Competing Financial Interests}\\
 C.E.L., R.L.R. and DeW.L.S. declare no competing financial interests
in the preparation of this work.

\vspace{0.5cm}
\noindent
\textbf{Figure captions}\\
\textbf{Figure 1.} Direct numerical simulations of a reconnection event at different time snapshots: 
$t=0$ interaction, $t=1$ reconnection, $t=2$ separation of tube strands. 
(a) Initially orthogonally-offset vortex tubes in a viscous fluid,
(b) quantized vortex tubes in superfluid helium, (c) magnetic flux tubes 
$\alpha$ and $\beta$ (centered on the spatial curves $C_\alpha$ and $C_\beta$) in 
magnetohydrodynamics. The top, central diagram shows a sketch at the reconnection 
site (in yellow), where the vortex strands become locally alligned in an anti-parallel 
fashion just before reconnection. Images adapted from \cite{ZM89}, \cite{ZCBB12} and 
\cite{P11}, respectively.

\noindent
\textbf{Figure 2.} (a) Reconnection of two oriented (polygonal) curves near a crossing does not change the 
writhe (since polygonal curves can approximate smooth curves arbitrarily closely, in this example 
we use polygonal curves). We assume that the curves remains almost co-planar at the crossing site, 
hence in all cases $Wr\approx-1$. Note the production of the `pigtail', due to the mutual 
cancellation of the anti-parallel strands. (b) Screen shots of the anti-parallel alignment and 
subsequent reconnection of two strands of a trefoil vortex knot from the experiment of Kleckner 
and Irvine${}^\textrm{\cite{KI13}}$ (reproduced with permission).  
(c) Smooth tracings of the screen shots, with the vortex overpasses made explicit. 
The apparent crossings at the bottom at each time sequence $t=0,1,2,3$ (red curve over blue curve) 
are the original overpasses of the same trefoil strands. The stage just after reconnection is shown 
in $t=3$. The directional writhe in each of the figures at $t=0,1,2,3$ is $+1$.
Compare this scenario with the idealized sketches above.

\noindent
\textbf{Figure 3.} Reconnection of polygonal curves $A$ and $B$: the intermediate $\theta$-curve
$(A\#B)^\ast$ has two coincident and oppositely oriented edges $a_n$ and $b_m$.

\noindent
\textbf{Figure 4.} (a) A flux tube $\gamma$ centered on the spatial curve $C_\gamma$. The ribbon 
$R_\gamma$ is formed by connecting $C_\gamma$ with one of the field lines in $\gamma$.
(b) Vortex lines and isosurface of vorticity (solid gray) under vortex tube reconnection. 
Note the bridge region formed by the re-organization of the weaker vorticity. 
From a direct numerical simulation of the Navier-Stokes equations.${}^\textrm{\cite{vRea12}}$

\end{document}